q
\documentstyle[pictex,12pt]{article}

\newcommand{\beq}{\begin{equation}}
\newcommand{\eeq}{\end{equation}}
\newcommand{\beqa}{\begin{eqnarray}}
\newcommand{\eeqa}{\end{eqnarray}}
\newcommand{\nn}{\nonumber}
\newcommand{\eq}[1]{(\ref{#1})}

\newcommand{\imtau}[1]{(2\pi\mbox{Im}\tau)_{#1}}

\newcommand{\alimit}{\stackrel{\alpha' \rightarrow 0}{\longrightarrow}}
\newcommand{\eqdef}{\stackrel{\mbox{def}}{=}}
\newcommand{\za}[1]{z_{\alpha_{#1}}}
\newcommand{\zb}[1]{z_{\beta_{#1}}}

\newcommand{\NP}[1]{ {\it Nucl.{}~Phys.} {\bf #1}}
\newcommand{\PL}[1]{ {\it Phys.{}~Lett.} {\bf #1}}

\newcommand{\PR}[1]{ {\it Phys.{}~Rev.} {\bf #1}}

\begin{document}
\topmargin 0pt
\oddsidemargin 1mm
\begin{titlepage}
\begin{flushright}
NBI-HE-97-39\\
hep-th/9709019\\
\end{flushright}
\setcounter{page}{0}

\vspace{20mm}
\begin{center}
{\Large Multiloop $\Phi^3$ Amplitudes from Bosonic String Theory.}
\vspace{20mm}

{\large Kaj 
Roland{}~\footnote{\,\,\,
roland@nbi.dk}}\\
{\em NORDITA \\
Blegdamsvej 17, DK-2100 Copenhagen, Denmark}\\
{\large Haru-Tada Sato   
\footnote{\,\,\,
sato@nbi.dk}}\\
{\em The Niels Bohr Institute \\
     Blegdamsvej 17, DK-2100 Copenhagen, Denmark}\\
\end{center}
\vspace{7mm}

\begin{abstract}
We show how the multiloop amplitudes of $\Phi^3$ theory (in the
worldline formulation of Schmidt and Schubert) are recovered from the
$N$-tachyon $(h+1)$-loop amplitude in bosonic string theory in the
$\alpha' \rightarrow 0$ limit, if one keeps the tachyon coupling
constant fixed. In this limit the integral over string moduli space
receives contributions only from those corners where the world-sheet
resembles a $\Phi^3$ particle diagram. Technical issues involved are a
proper choice of local world-sheet coordinates, needed to take the
string amplitudes off-shell, and a formal procedure for introducing a
free mass parameter $M^2$ instead of the tachyonic value $-4/\alpha'$.
\end{abstract}

\vspace{1cm}

\end{titlepage}
\newpage
\renewcommand{\thefootnote}{\arabic{footnote}}
\section{Introduction}
\setcounter{equation}{0}
In the limit of vanishing inverse string tension, $\alpha' \rightarrow
0$, any string theory reduces to an effective field theory. Since in
this limit the string shrinks to a particle-like object,
it is intuitively clear that the Polyakov path integral over
possible world-sheet histories, which defines the string scattering
amplitudes, must reduce to a path integral over world-line histories. 
Thus, by considering the $\alpha' \rightarrow 0$ limit of string
theory amplitudes one arrives automatically at a world-line
formulation of the particle theory amplitudes associated with the
low-energy effective field theory. Sometimes, the world-line
formulation thus obtained offers a considerable improvement in
computational efficiency over ordinary Feynman diagram techniques. The
Bern-Kosower rules for one-loop Yang-Mills theory~\cite{BK,BDK}
constituted the
first, and most spectacular, example of this.

World-line formulations of particle theory may also be obtained
directly from field theory~\cite{wline,mixed,SSphi,SSqed,sumino}. 
This approach has the obvious advantage of
bypassing the complexities of string theory altogether. This is
particularly appealing in the case of multiloop amplitudes. On the
other hand, some aspects of the Bern-Kosower rules (such as the subtle
combination of gauge choices and the integration-by-parts procedure)
would have been quite hard to
discover without the help of string theory, and it is therefore
natural to expect that the $\alpha' \rightarrow 0$ limit of string
theory can be helpful also in formulating multiloop extensions of the
Bern-Kosower rules.

However, the $\alpha' \rightarrow 0$ limit of string multiloop gluon amplitudes
is technically very intricate; many of the complications arise from
the presence of contact terms. This
problem can be avoided by considering instead (as
a toy model) the case of multiloop $\Phi^3$ theory, 
as has been advocated already by
Di Vecchia et al~\cite{divecchia}: 
By taking the $\alpha' \rightarrow 0$ limit of
$N$-tachyon $(h+1)$-loop amplitudes, keeping the $3$-tachyon coupling
constant $g$ fixed, one expects to obtain the $N$-point $(h+1)$-loop
amplitudes of $\Phi^3$-theory, with a mass 
\beq
M^2 = -\frac{4}{\alpha'} \rightarrow - \infty \ .
\eeq
This is due to the fact that in string theory any contribution to the
tree-level $N$-tachyon amplitude is proportional to $N-2$ powers of
the string coupling constant and hence to $N-2$ powers of the tachyon
coupling constant $g$. Therefore, on dimensional grounds alone, any
would-be $\Phi^N$ coupling ($N \geq 4$) in the effective tachyon field
theory must be proportional to
\beq
g^{N-2} (\alpha')^{N-3} \ , 
\eeq
and hence vanishing in the $\alpha' \rightarrow 0$ and $g$ fixed limit.

As we shall see, surviving contributions to the $N$-tachyon
$(h+1)$-loop amplitudes in the $\alpha' \rightarrow 0$ and $g$ fixed
limit come only from infinitesimal regions around the singular
points on the boundary of moduli space, where the string
world-sheet degenerates into a $\Phi^3$ particle diagram.
In such a region (also called a {\em $\Phi^3$-like corner of moduli
space}) the world-sheet consists of cylinders, 
very long in units of the diameter, joined
together at $N+2h$ vertices. These corners of moduli space were
studied extensively in ref.~\cite{RS}, where it was shown how to map the
world-sheet moduli into the Schwinger proper times (SPTs) of the
corresponding $\Phi^3$ particle diagram.

In the Schwinger Proper Time (SPT) parametrization, the negative sign
of $M^2$ gives rise to factors which are exponentially enhanced, rather than
exponentially suppressed,
\beq
\exp \{ - M^2 \tau \} \ = \ \exp \{ \frac{4}{\alpha'} \tau \} \ .
\label{expfactor} 
\eeq
Therefore, the integral over the SPT $\tau$ diverges for large values
of $\tau$. In the limit
$\alpha' \rightarrow 0$ for fixed $\tau$ 
even the {\em integrand} becomes ill-defined.
However, if one formally replaces the $4/\alpha'$ appearing in the
exponent by $-M^2$ and takes
$M^2$ to be a free positive parameter, then one obtains the multiloop
amplitudes of $\Phi^3$ theory with an arbitrary mass. Since the string
theory starting point is that $\alpha' M^2 = -4$, this procedure
requires that one ``knows'' when to interprete the number $-4$ as
$\alpha' M^2$ and when not to do so. This potential ambiguity is avoided
because of the fact that, in $\Phi^3$-theory, the particle mass enters
the {\em off-shell} amplitudes {\em only} through the propagators,
i.e. in the SPT parametrization through factors such as
\eq{expfactor}, which are easily identified. If the amplitudes are
on-shell, on the other hand, the particle mass enters also through the
mass-shell condition $p_i^2=-M^2$ for the external states, $i=1,\ldots,N$.

It is well known that, when one considers off-shell string amplitudes,
i.e. relaxes the condition $\alpha' p_i^2 = 4$ for the $i$'th
external tachyon state ($i=1,\ldots, N$), the integrated vertex operator 
\beq
\int \mbox{d}^2 z  \ e^{i p_i \cdot X (z,\bar{z})} 
\eeq
develops a dependence on the choice of holomorphic coordinate, $z$.
Let $w_i$ denote the coordinate chosen for the $i$'th vertex
operator. This coordinate may depend in a complicated way 
on the moduli of the string world-sheet.
In ref.~\cite{RS} a prescription was given for $w_i$ in the $\Phi^3$-like
corners of moduli space.

In the present paper we show how this prescription for the local
coordinates, together with the above procedure for introducing a free
mass parameter $M^2$, gives rise, in the $\alpha' \rightarrow 0$ and $g$
fixed limit, to a worldline formula for $N$-point
$(h+1)$-loop $\Phi^3$ amplitudes that correctly reproduces the SPT
integrand appearing in the formulae
obtained from field theory in refs.~\cite{SSphi,RS}, up to an overall
normalization constant. 

The paper is organized as follows: In section 2 we review the worldline
formulation of $\Phi^3$-theory. In section 3 we consider the string
theory formalism and argue that only $\Phi^3$-like corners of moduli
space contribute to the $N$-tachyon amplitude in the $\alpha'
\rightarrow 0$ and $g$ fixed limit. In section 4 we study a large
class of
$\Phi^3$-like corners in detail and show how the string modular
integrand reduces to the SPT integrand of the particle worldline
formula. Section 5
contains some remarks on the region of integration. Finally, in Appendix
A we show how the symmetric SPT parametrization available in the
two-loop case is contained in our general discussion,  
and in Appendix B we provide certain technical
details needed for the comparison of the string and particle
formulae.

\section{Particle Theory Formalism}
\setcounter{equation}{0}
The world-line formulation of $\Phi^3$ theory has been developed by
Schmidt and Schubert~\cite{SSphi}. In their formalism it is possible to sum
large classes of $N$-point $(h+1)$-loop Feynman diagrams by
means of a single SPT integral formula. 

Included is any diagram that can be obtained by inserting
$N$ external legs on a {\em Schmidt-Schubert type vacuum diagram}, that
is, on a vacuum diagram built by connecting $2h$ points inserted on a 
circle with $h$ {\it internal propagators} (see Fig. 1). We let ${\cal
C}_{N_0,N_1,\ldots,N_h}^{(h+1)}$ denote the class of $\Phi^3$-diagrams where
$N_0$ external legs (with incoming momenta $p_n^{(0)},
n=1,\ldots,N_0$) are inserted on the circle (also known as {\it the
fundamental loop}) and $N_i$ external legs (with momenta $p_n^{(i)},
n=1,\ldots,N_i$) are inserted on the $i$'th internal
propagator, $i=1,\ldots,h$.
\begin{figure}
\begin{center}
\input{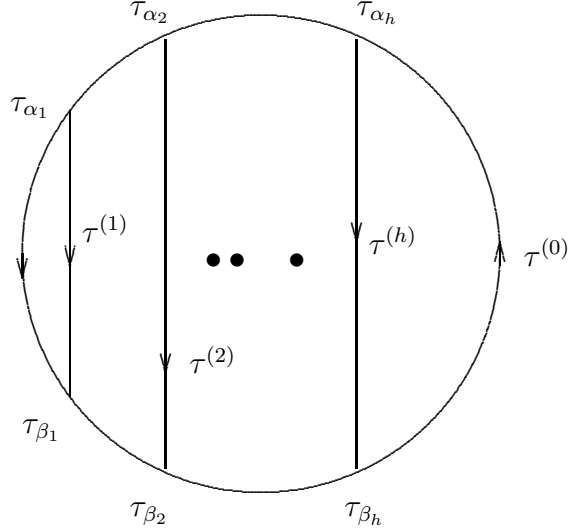}
\end{center}
\caption{A Schmidt-Schubert type vacuum diagram. The arrows indicate
the direction of increasing Schwinger Proper Time. All possible
orderings of the $2h$ vertices are allowed.}
\end{figure}

Along the fundamental loop we define a SPT $\tau^{(0)} \in [0,T]$ and
along the $i$'th internal propagator we introduce a SPT $\tau^{(i)}
\in [0,\bar{T}_i]$, $i=1,\ldots, h$. The points where
the $i$'th internal propagator is joined to the fundamental loop 
are denoted by $\tau^{(0)}=\tau_{\alpha_i}$, corresponding to
$\tau^{(i)}=0$, and by $\tau^{(0)}=\tau_{\beta_i}$, corresponding to
$\tau^{(i)}=\bar{T}_i$. The $N_0$ legs inserted on the fundamental
loop have SPTs $\tau_n^{(0)} \in [0;T]$, $n = 1, \ldots, N_0$, 
and the $N_i$ external legs on the $i$'th internal propagator have 
SPTs $\tau_n^{(i)} \in [0;\bar{T}_i]$, $n = 1, \ldots, N_i$.

The corresponding amplitude,
equal to the sum of all the diagrams in ${\cal
C}_{N_0,\ldots,N_h}^{(h+1)}$, is given by
\begin{eqnarray}
\Gamma_{N_0, \ldots, N_h}^{(h+1)}& = & {\cal N} \
g^{N+2h}\int_0^{\infty}{\mbox{d}T\over T} T^{-D/2}  
\ \cdot \prod_{i=1}^h \int_0^{\infty} \mbox{d}
{\bar T}_i
\int_0^T \mbox{d} \tau_{\alpha_i} \int_0^T \mbox{d}
\tau_{\beta_i}  \nonumber \\
& & \times \prod_{n=1}^{N_0} \int_0^T \mbox{d}\tau_n^{(0)} \cdot
\prod_{i=1}^h \prod_{n=1}^{N_i} \int_0^{\bar{T}_i} \mbox{d}
\tau_n^{(i)} \cdot \  \ (\mbox{det}A)^{-D/2}\ 
\exp\left[-M^2(T+\sum_{i=1}^h{\bar T}_i)\right]  \nn\\
& & \times \ \exp\left[{1\over2}\sum_{i,j=0}^h\sum_{n=1}^{N_i}
    \sum_{n'=1}^{N_j}p_n^{(i)} p_{n'}^{(j)} 
G_{ij}^{(h)}(\tau_n^{(i)},\tau_{n'}^{(j)})\right] \ ,  \label{ptamp}
\end{eqnarray}
where ${\cal N}$ is a normalization constant (which we do not try to
fix), $g$ is the $\Phi^3$ coupling constant, and 
\beq
N=\sum_{i=0}^h N_i 
\eeq
is the total number of external legs.

The various Green functions appearing in \eq{ptamp} are given
by~\cite{SSphi,RS} 
\beqa
G_{00}^{(h)} (\tau_1^{(0)},\tau_2^{(0)}) & = &
G_B(\tau_1^{(0)},\tau_2^{(0)}) - \sum_{k,l=1}^h
X_k(\tau_1^{(0)},\tau_2^{(0)}) A^{-1}_{kl}
X_l(\tau_1^{(0)},\tau_2^{(0)}) \ , \label{gh00} \\
G_{ii}^{(h)}(\tau_1^{(i)},\tau_2^{(i)})  & = &
\vert\tau_1^{(i)}-\tau_2^{(i)}\vert - 
(\tau_1^{(i)}-\tau_{2}^{(i)})^2 A^{-1}_{ii} \ \ , \  i = 1, \ldots, h
\ , \label{ghii} \\
G_{i0}^{(h)}(\tau_1^{(i)},\tau_2^{(0)}) & = &  
\tau_1^{(i)}+ G_B(\tau_{\alpha_i},\tau_2^{(0)})  \label{ghi0}  \\
& & - \sum_{k,l=1}^h 
[ -\tau_1^{(i)}\delta_{ik} +X_k(\tau_{\alpha_i},\tau_2^{(0)})] 
A^{-1}_{kl}
[ -\tau_1^{(i)}\delta_{il} +X_l(\tau_{\alpha_i},\tau_2^{(0)})]
\ , \nn \\
& & \hspace{6cm} i = 1, \ldots , h \ ,  \nn 
\eeqa
\beqa
\lefteqn{G_{ij}^{(h)}(\tau_1^{(i)},\tau_2^{(j)}) \  = \  
\tau_1^{(i)}+\tau_2^{(j)}+G_B(\tau_{\alpha_i},\tau_{\alpha_j})} 
\label{ghij} \\
& &  - \sum_{k,l=1}^h
[-\tau_1^{(i)}\delta_{ik} + \tau_2^{(j)}\delta_{jk}
  +X_k(\tau_{\alpha_i},\tau_{\alpha_j})] A^{-1}_{kl} 
[-\tau_1^{(i)}\delta_{il} + \tau_2^{(j)}\delta_{jl}
  +X_l(\tau_{\alpha_i},\tau_{\alpha_j})]\ , \nonumber \\
& & \hspace{9cm} i,j = 1, \ldots, h , \ , i \neq j \ , \nn 
\eeqa
where $A$ is the $h\times h$ matrix defined by  
\beq
 A_{ij}\equiv{\bar T}_i\delta_{ij} -
         X(\tau_{\alpha_i},\tau_{\beta_i};
           \tau_{\alpha_j},\tau_{\beta_j}) \ ,           
    \label{matA}
\eeq
\beq
X(\tau_a,\tau_b;\tau_c,\tau_d)\equiv{1\over2}
        \left[ G_B(\tau_a,\tau_c)-G_B(\tau_a,\tau_d)
       -G_B(\tau_b,\tau_c)+G_B(\tau_b,\tau_d)\right] \ , 
    \label{Xabcd} 
\eeq
$G_B$ is the bosonic Green function on a loop of SPT length
$T$~\cite{wline},
\beq
G_B (\tau_1,\tau_2) = \vert \tau_1 - \tau_2 \vert - 
\frac{(\tau_1-\tau_2)^2}{T} \ ,                   \label{GB}
\eeq
and we introduced the shorthand notation
\beq
X_j(\tau_a,\tau_b) \equiv 
X(\tau_a,\tau_b;\tau_{\alpha_j},\tau_{\beta_j}) \ .  
\label{short2}
\eeq
It may be noted that the integrand in eq.~\eq{ptamp} remains invariant
if one adds the same constant, $\Delta \tau$, to {\em all} the SPT
variables defined on the fundamental loop, i.e. the variables
$\tau_{\alpha_i}$, $\tau_{\beta_i}$, $i=1,\ldots, h$ and
$\tau_n^{(0)}$, $n=1,\ldots, N_0$. As argued in detail in Appendix B
one may use this symmetry to 
remove all dependence on, say, $\tau_{\alpha_1}$. The resulting empty
integration can then be trivially performed to give just a factor of
$T$ and one obtains
\begin{eqnarray}
\Gamma_{N_0, \ldots, N_h}^{(h+1)}& = & {\cal N} \
g^{N+2h}\int_0^{\infty} \mbox{d}T T^{-D/2}  
\ \cdot \prod_{i=1}^h \int_0^{\infty} \mbox{d}
{\bar T}_i \cdot \int_0^T \mbox{d} \tau_{\beta_1}
\cdot \prod_{i=2}^h \int_0^T \mbox{d} \tau_{\alpha_i} \int_0^T \mbox{d}
\tau_{\beta_i}  \nonumber \\
& & \times \prod_{n=1}^{N_0} \int_0^T \mbox{d}\tau_n^{(0)} \cdot
\prod_{i=1}^h \prod_{n=1}^{N_i} \int_0^{\bar{T}_i} \mbox{d}
\tau_n^{(i)} \cdot \  \ (\mbox{det}A)^{-D/2}\ 
\exp\left[-M^2(T+\sum_{i=1}^h{\bar T}_i)\right]  \nn\\
& & \times \ \exp\left[{1\over2}\sum_{i,j=0}^h \sum_{n=1}^{N_i}
    \sum_{n'=1}^{N_j}p_n^{(i)} p_{n'}^{(j)} 
G_{ij}^{(h)}(\tau_n^{(i)},\tau_{n'}^{(j)})\right] \ ,  \label{ptamptwo}
\end{eqnarray}
where now everywhere $\tau_{\alpha_1} = 0$.
This is the formula that we will now reproduce from string theory.

\section{String Theory Formalism}
\setcounter{equation}{0}
In string theory the entire connected $N$-tachyon $(h+1)$-loop
amplitude is given by the ``master formula''
\beqa
\lefteqn{T^{(h+1)} (p_1, \ldots p_N) = }  \nonumber \\
& & C_{h+1} \left( \frac{\kappa}{\pi} \right)^N
\ \int \mbox{d}^2 {\cal M} \ (\mbox{det}\,2\pi\mbox{Im}\tau)^{-D/2} \
\exp\left[{\alpha'\over2}\sum_{i \neq j=1}^N p_i\cdot p_j 
\, G^{(h)}_{\rm str}(z_i,z_j)\right] \ ,              \label{stramp}
\eeqa
where $C_{h+1}$ is the normalization of the $(h+1)$-loop vacuum
amplitude, given in terms of the gravitational coupling, $\kappa$,
by~\cite{unitarity} 
\beq
C_{h+1} = \left( \frac{2\kappa^2}{\alpha'} \right)^h \left(
\frac{1}{2\pi} \right)^{D(h+1)/2 + 3h} \left( \alpha' \right)^{-2+
(h+1)(4-D)/2}  \ , \label{vacuumconst}
\eeq
and the $N$ powers of $\kappa / \pi$ are due to the normalization of
the vertex operators~\cite{normherm}. The integral is over one copy of 
$N$-punctured genus $(h+1)$ moduli space. 

Each point in this space corresponds to an $N$-punctured genus $(h+1)$
Riemann surface. In the Schottky parametrization~\cite{Schottky} 
the Riemann surface is
identified with the sphere, which has a global complex coordinate $z$
(defined up to an overall projective transformation),
modulo the action of the discrete group of projective 
transformations generated by $h+1$ generators; moduli space is
parametrized by the multiplier $k_{\mu}$ and the fixed points
$z_{\alpha_{\mu}}$ and $z_{\beta_{\mu}}$ of these
generators ($\mu=0,1,\ldots, h$) and by the Koba-Nielsen variables
$z_i$, $i=1,\ldots,N$, 
specifying the positions of the vertex operators. 

The SPT length of the $\mu$'th
closed loop is given by~\cite{marco,Rolandb}
\beq
T_{\mu} = - \frac{\alpha'}{2} \ln | k_{\mu} | \ , \label{Tmudef}
\eeq
and in the limit where we take $\alpha' \rightarrow 0$, $T_{\mu}$ will
vanish unless, at the same time, $k_{\mu}$ becomes very small. 
It is important to
notice that even in theories such as Yang-Mills theory, where 
contact term contributions can arise from internal propagators of
vanishing SPT length, the SPT length $T_{\mu}$ of an {\em
entire closed loop} must remain finite if
all $h+1$ loops are to survive in the $\alpha' \rightarrow 0$ limit.
For this reason we may always
expand in powers of $k_{\mu}$ and $\bar{k}_{\mu}$. 
As explained in ref.~\cite{unitarity} it follows directly
from the sewing procedure that the contribution to
the modular integrand due to tachyons circulating in the $\mu$'th loop is
obtained as the leading term in this expansion, proportional to
$\mbox{d}^2 k_{\mu} / |k_{\mu}|^4$. Ignoring all contributions of
higher order in $k_{\mu}$ and $\bar{k}_{\mu}$ (which account for the
exchange of massless states and of states with positive $M^2$) one
finds~\cite{plb} 
\beq
\mbox{d}^2 {\cal M}={1\over \mbox{d}^2V_{abc}} \cdot \prod_{\mu=0}^h 
{ \mbox{d}^2k_\mu \over |k_\mu|^4} { \mbox{d}^2z_{\alpha_\mu}
\mbox{d}^2z_{\beta_\mu} \over  
|z_{\alpha_\mu}-z_{\beta_\mu}|^4} \cdot 
\prod_{i=1}^N {\mbox{d}^2z_i\over |V'_i(0)|^2}  \ ,            \label{measure}
\eeq
where
\beq
\mbox{d}^2 V_{abc}=
\frac{\mbox{d}^2 z_a \mbox{d}^2 z_b \mbox{d}^2 z_c}{ \vert
(z_a-z_b)(z_a-z_c)(z_b-z_c) \vert^2} \ , \label{pvolume}
\eeq
and it is understood that projective invariance is used to fix three
of the $2h+ N+2$ points $\{ z_{\alpha_{\mu}}, z_{\beta_{\mu}}, z_i \}$
at definite values $z_a$, $z_b$ and $z_c$.

By definition the factor
\beq
V_i'(0) = \left. \frac{\mbox{d} z}{\mbox{d} w_i} \right|_{w_i=0} 
\label{loccoord}
\eeq
relates the global coordinate $z$, which assumes the value $z_i$ at
the point where the $i$'th vertex operator is inserted, to
the local coordinate $w_i$. (It is conventional to take $w_i=0$ at the
point $z=z_i$.)
For on-shell tachyons ($\alpha' p^2 = +4$)
the factors \eq{loccoord} in $\mbox{d}^2 {\cal M}$ cancel
against similar factors appearing in the bosonic Green
function~\cite{Martinec},
\beq
G^{(h)}_{\rm str}(z_1,z_2)=\ln\left\vert
    {E(z_1,z_2) \over (V_1'(0) V_2'(0))^{1/2}} \right\vert  
-{1\over2}\sum_{\mu,\nu=0}^h\Omega_\mu\imtau{\mu\nu}\Omega_\nu\ , 
\label{sgreen} 
\eeq
where, to leading order in the multipliers, the prime form, period
matrix and Abelian integrals are given by~\cite{Schottky}
\begin{eqnarray}
E(z_1,z_2) & = & z_1 - z_2 \ , 
\label{prime} \\
(2\pi\mbox{Im}\tau)_{\mu\nu} & = & 
-\delta_{\mu\nu}\ln\vert{k_\mu}\vert 
-(1-\delta_{\mu\nu})\ln  \left\vert
{z_{\alpha_\mu}-z_{\alpha_\nu}\over 
z_{\alpha_\mu}-z_{\beta_\nu}}
{z_{\beta_\mu}-z_{\beta_\nu}\over 
z_{\beta_\mu}-z_{\alpha_\nu}}   
                          \right\vert  \ , 
\label{period} \\                     
\Omega_\mu  & = & \ln \left\vert
{z_2-z_{\beta_\mu}\over z_1-z_{\beta_\mu}}
{z_1-z_{\alpha_\mu}\over z_2-z_{\alpha_\mu}} \right\vert 
 \ . \label{omega}
\end{eqnarray}
For off-shell tachyons the amplitude depends on the choice of
$|V_i'(0)|$, which we choose in accordance with the prescription of
ref.~\cite{RS}: In any $\Phi^3$-like corner of moduli space each vertex
operator insertion point $z_i$ represents an external leg inserted
on some very long cylinder,
and we choose
\beq
|V_i'(0)|^{-1} = | \omega(z_i) | \ , \label{choice}
\eeq
where $\omega$ is any one-differential which is holomorphic throughout
the cylinder and has period $2\pi i$ along any contour encircling the
cylinder (if we change the orientation of the contour, $V_i'(0)$
changes sign; but that is not important since we are only interested 
in $|V_i'(0)|$). As
one approaches the singular point on the boundary of moduli space,
where the world-sheet degenerates into the $\Phi^3$-diagram, the
cylinder becomes infinitely long (in units of the width) and $\omega$
becomes unique.

If we start by evaluating eq.~\eq{stramp} in the case $h+1=0$ and
$N=3$, we find the precise relation between the gravitational coupling
$\kappa$ and the tachyon coupling constant, $g$:
\beq
T^{(0)} (p_1,p_2,p_3) = {4\kappa \over \alpha'} \eqdef g \ . \label{gdef}
\eeq
The expression \eq{gdef} for $T^{(0)} (p_1,p_2,p_3)$ holds both on and
off the mass-shell, provided that one chooses 
\beq
|V_i'(0)|^{-1} = \left|
\frac{(z_{i-1}-z_{i+1})}{(z_i-z_{i-1})(z_i-z_{i+1})} \right| \ ,
\label{lovelace}
\eeq
where we adopted a cyclic notation
(i.e. $z_{i+3} = z_i$). This is in fact the well-known Lovelace
choice~\cite{Lovelace}. 
It can be obtained from our general prescription \eq{choice} in the
following sense: The world-sheet (there is only one, since there are no moduli
for a three-punctured sphere) consists of three semi-infinite
cylinders joined together at a vertex. The string state propagating
along the $i$'th cylinder is represented as a vertex operator inserted
at the point $z_i$. From the point of view of this vertex operator,
the world-sheet looks like an infinite cylinder, represented as the
complex sphere with the points $z=z_{i-1}$ and $z=z_{i+1}$
removed. The unique holomorphic one-differential of period $2\pi i$ is
given by
\beq
\omega(z) = \frac{z_{i-1}-z_{i+1}}{(z-z_{i-1})(z-z_{i+1})} \ , 
\eeq
from which eq.~\eq{lovelace} is immediately obtained, if we take
$z=z_i$ as in eq.~\eq{choice}.

Using eqs.~\eq{vacuumconst} and \eq{gdef} we may rewrite the
overall numerical constant appearing in the ``master formula''
\eq{stramp} as follows
\beqa
\lefteqn{ C_{h+1} \left( \frac{\kappa}{\pi} \right)^N = }
\label{overallconstant} \\
& & g^{N+2h} \left( \alpha' \right)^{N+3h-(h+1)D/2} \left(
\frac{1}{2\pi} \right)^{N+D(h+1)/2+3h} \left( \frac{1}{2}
\right)^{3h+N} \ . \nn
\eeqa 
To properly count the powers of $\alpha'$ one must recall the relation
\eq{Tmudef} between the diagonal elements of the matrix $(2\pi
\mbox{Im} \tau)_{\mu \nu}$ and the SPT lengths $T_{\mu}$,
$\mu=0,1,\ldots,h$, of the $h+1$ closed
loops. Since $T_{\mu}$ must remain finite in the $\alpha' \rightarrow
0$ limit, $(\mbox{det} 2\pi
\mbox{Im} \tau)^{-D/2}$ scales like $(\alpha')^{(h+1)D/2}$;
therefore, for fixed $T_{\mu}$
\beq
C_{h+1} \left( \frac{\kappa}{\pi} \right)^N \left( \det 2\pi \mbox{Im}
\tau \right)^{-D/2} \propto (\alpha')^{N+3h} \ . \label{alphaprimecount}
\eeq
In conclusion, we obtain a finite amplitude from the string ``master
formula'' \eq{stramp} in the $\alpha' \rightarrow 0$ and $g$ fixed
limit if and only if the measure scales like
\beq
\mbox{d}^2 {\cal M} \propto (\alpha')^{-(N+3h)} \
. \label{measurescale}
\eeq
Now, whenever we change integration variables in $\mbox{d}^2 {\cal M}$
from a complex modular parameter to a SPT {\em which is kept finite in
the $\alpha' \rightarrow 0$ limit} we obtain a factor of
$1/\alpha'$. As an example of this we may consider the change of variables 
from $k_{\mu}= |k_{\mu}| e^{i\phi_{k_{\mu}}}$ to $T_{\mu}$:
\beqa
\frac{\mbox{d}^2 k_{\mu}}{|k_{\mu}|^4} & = & \frac{\mbox{d}
|k_{\mu}|}{|k_{\mu}|} |k_{\mu}|^{-2} \mbox{d} \phi_{k_{\mu}} \ = \
\frac{2}{\alpha'} \mbox{d} T_{\mu} e^{-2 \log |k_{\mu}|} \mbox{d}
\phi_{k_{\mu}} \nonumber \\
& = & \frac{2}{\alpha'} \mbox{d} T_{\mu} \mbox{d}
\phi_{k_{\mu}} e^{+\frac{4}{\alpha'} T_{\mu}} \ = \
\frac{2}{\alpha'} \mbox{d} T_{\mu} \mbox{d}
\phi_{k_{\mu}} e^{-M^2 T_{\mu}} \nonumber \ ,
\eeqa
where we introduced the tachyon mass $M$ like in eq.~\eq{expfactor}.

We see that in order to obtain the scaling behaviour \eq{measurescale}
we {\em have} to trade {\em all} the $3h+N$ complex moduli for a {\em
finite} SPT, plus a phase. In other words, we have to 
approach the various $\Phi^3$-like corners of moduli space.  In
each $\Phi^3$-like corner of moduli space there is a one-to-one map of
the $3h+N$ complex string moduli into the SPTs of the corresponding
$\Phi^3$ diagram (plus $3h+N$ phases that just give rise to empty
integrations), and when one takes the limit $\alpha' \rightarrow 0$,
keeping all the $3h+N$ SPTs fixed, one approaches the singular point on the
boundary of moduli space where the world-sheet degenerates into the
$\Phi^3$ particle diagram in question~\cite{RS}.

Thus, instead of a single integral over all of moduli space, as in
eq.~\eq{stramp}, we obtain a sum of contributions from all
$\Phi^3$-like corners, each contribution being on the form of a $3h+N$
real-dimensional SPT integral.

\section{Particle Theory from String Theory}
\setcounter{equation}{0}
Having made these general observations we would now like to consider
in detail how the particle formula \eq{ptamptwo} can be obtained from
the string ``master formula'' \eq{stramp} precisely by restricting
ourselves to corners of $N$-punctured genus $(h+1)$
moduli space, where the world sheet resembles a $\Phi^3$-diagram constructed
by inserting $N$ external legs on a Schmidt-Schubert type vacuum diagram.

It was shown in ref.~\cite{RS} how, in any such corner of $N$-punctured genus
$(h+1)$ moduli space, where $z_i$ describes an external line inserted
on the $\mu$'th internal propagator with SPT $\tau_n^{(\mu)}$ and
$z_j$ a line inserted on the $\nu$'th internal propagator with SPT
$\tau_{n'}^{(\nu)}$ ($\mu,\nu = 0,1,\ldots,h$)~\footnote{The $0$'th
internal propagator is the fundamental loop.} one finds
\beq
G^{(h)}_{\rm str}(z_i,z_j) \quad\alimit\quad 
{1\over\alpha'}G_{\mu \nu}^{(h)}(\tau_n^{(\mu)},\tau_{n'}^{(\nu)}) \,
\eeq
in the limit $\alpha' \rightarrow 0$, when all the SPTs are kept
fixed, provided that the
coordinate-dependent factors $|V_i'(0)|$ are chosen in accordance with
the prescription already described. 

Thus, with the proper choice of $V_i'(0)$ we find
\beq
\exp\left[{\alpha'\over2}\sum_{i \neq j=1}^N p_i\cdot p_j 
G^{(h)}_{\rm str}(z_i,z_j)\right]  \ \alimit \ 
\exp\left[{1\over2}\sum_{i,j=0}^h\sum_{n=1}^{N_i}\sum_{n'=1}^{N_j}
p_n^{(i)} p_{n'}^{(j)} 
G_{ij}^{(h)}(\tau_n^{(i)},\tau_{n'}^{(j)})\right] \, . \label{exponent}
\eeq
(Strictly speaking one obtains the exponent of the particle formula
\eq{ptamptwo} {\em except} for the diagonal terms, where $i=j$ and
$n=n'$. But these vanish anyway, due to the explicit form of the Green
functions \eq{gh00} and \eq{ghii}.)
It remains to investigate the behaviour of the period matrix
determinant and of the
integration measure. 

We choose to represent the fundamental loop 
by the $0$'th string loop and we use
projective invariance on the sphere to fix
\beq
z_{\alpha_0}=\infty \ \ \ \mbox{and} \ \ \ z_{\beta_0}=0 \ .
\eeq
If we introduce the notation
\beqa
\imtau{00} &=& \Delta \ , \label{periodmatrix} \\
\imtau{i0} &=& \imtau{0i} = \ln\left|{\za{i}\over\zb{i}}\right| = x_i
\ \ \ \mbox{for} \ \ i = 1, \ldots , h \ , 
\eeqa
the determinant is obtained as follows:
\beq
\mbox{det}(2\pi\mbox{Im}\tau) = 
\left|\begin{array}{llll} 
\Delta   & x_1  &x_2\,\,\cdots  &x_h\\
   x_1   & { }  &{}           &{} \\
   x_2   & { }  &\imtau{ij}   &{} \\
\dot{:}  & { }  &{}           &{} \\
   x_h   & { }  &{}           &{} \end{array}\right|\,.
\eeq
Multiply the $i$-th row by $\Delta$, and then 
subtract the $0$-th row multiplied by $x_i$ from the $i$-th row. 
Repeating this for $i=1,2,\cdots, h$, we obtain 
\beq
\mbox{det}(2\pi\mbox{Im}\tau) = 
\Delta^{1-h} \det \left[ \Delta A^{\rm str}_{ij} \right] \ ,
\eeq
where the $h \times h$ matrix $A_{ij}^{\rm str}$ is given by
\beq
A^{\rm str}_{ij} = \imtau{ij}-x_ix_j\Delta^{-1} \ . 
\eeq
This matrix is closely related to the matrix $A$ appearing in the
particle formula \eq{ptamp}. Indeed, as was shown in ref.~\cite{RS}, if we
take the limit $\alpha' \rightarrow 0$,
keeping the SPTs fixed, we have
\beq
A_{ij}^{\rm str} \quad \alimit \quad \frac{2}{\alpha'} A_{ij} 
\eeq
and thus
\beqa
\mbox{det} (2 \pi \mbox{Im} \tau) &&= \Delta\ \mbox{det}A^{\rm str}_{ij} \nn \\
  & &\alimit  \left({2\over\alpha'}\right)^{h+1}T \ \mbox{det}A \ ,
\label{determinantformula} 
\eeqa 
where we also used the relation 
\beq
T = \frac{\alpha'}{2} \Delta
\eeq
between $\Delta$ and the SPT $T$ of the fundamental loop, 
q.v. eqs. \eq{Tmudef}, \eq{period} and \eq{periodmatrix}. 

Consider finally the integration measure. 

Fixing $z_a=\za{0}=\infty$, $z_b=\zb{0}=0$ and $z_c=z_{\alpha_1}=1$ in 
\eq{pvolume}, the integration measure \eq{measure} becomes 
\beq
\mbox{d}^2{\cal M} = \mbox{d}^2{\cal M}_1 \mbox{d}^2{\cal M}_2 \ , 
\eeq
where
\beq
\mbox{d}^2{\cal M}_1 = {\mbox{d}^2k_0 \over |k_0|^4} \cdot
\prod_{i=1}^h {\mbox{d}^2k_i \over  |k_i|^4} \cdot  {|z_{\alpha_1}|^2
\mbox{d}^2 z_{\beta_1} \over |z_{\alpha_1}-z_{\beta_1}|^4} \cdot
\prod_{i=2}^h {\mbox{d}^2z_{\alpha_i} \mbox{d}^2z_{\beta_i} 
\over |z_{\alpha_i}-z_{\beta_i}|^4} \ ,  \label{dm1}
\eeq
\beq
\mbox{d}^2{\cal M}_2 = \prod_{i=1}^N {\mbox{d}^2z_i\over |V'_i(0)|^2} \ .        \label{dm2}
\eeq
Let us consider piece by piece. The relations between SPTs and
string moduli for the Schmidt-Schubert type vacuum diagram are given
by~\cite{RS} 
\beqa
\tau_{\alpha_i} &=& -{\alpha'\over2}\ln|\za{i}| \quad 
\mbox{for} \quad i=1,\cdots,h,                  \label{tauai} \\
\tau_{\beta_i}  &=& -{\alpha'\over2}\ln|\zb{i}| \quad
\mbox{for} \quad i=1,\cdots,h, \nn \\
T & = & -{\alpha'\over2}\ln|k_0| \nn \\
T_i  &=& -{\alpha'\over2}\ln|k_i| \quad 
\mbox{for} \quad i=1,\cdots,h \ .        \nn 
\eeqa
Here $T_i$ is the SPT of the entire $i$'th loop. The SPT of the $i$'th
internal propagator is given by
\beqa
{\bar T}_i &=& T_i - |\tau_{\alpha_i}-\tau_{\beta_i}| \ = \
-{\alpha'\over2}\ln\left| 
{k_i(\za{i} -\zb{i})^2 \over \za{i}\zb{i}} \right| 
\quad \mbox{for} \quad i=1,\cdots,h,  \label{Tintprop}
\eeqa
where one should keep in mind that $\tau_{\alpha_i} > \tau_{\beta_i}$
($\tau_{\alpha_i} < \tau_{\beta_i}$) corresponds to $|z_{\alpha_i}| \ll
|z_{\beta_i}|$ ($|z_{\alpha_i}| \gg |z_{\beta_i}|$) in the $\alpha'
\rightarrow 0$ limit. Using these equations
we find 
\beqa
\mbox{d}^2{\cal M}_1 & \alimit & \left( {4\pi\over\alpha'}\right)^{3h} 
\mbox{d}T \left(\prod_{i=1}^h \mbox{d}\bar{T}_i \right)
\mbox{d} \tau_{\beta_1} \left( \prod_{i=2}^h
\mbox{d}\tau_{\alpha_i} \mbox{d}\tau_{\beta_i} \right) 
\exp\left[ {4\over\alpha'}(T+\sum_{i=1}^h{\bar T}_i)\right] \ .
\label{dmone} 
\eeqa
Since at this point the integrand no longer depends on the {\em
phases} of the modular parameters, the corresponding integrations
become trivial and produce the factor $(2\pi)^{3h}$ appearing above.

Now let us consider $\mbox{d}^2{\cal M}_2$. 
If $z_i$ describes a leg inserted on the fundamental loop, then, as shown in
ref.~\cite{RS}, the corresponding SPT $\tau^{(0)}$ and the local coordinate
factor $V_i'(0)$ are given by
\beq
\tau^{(0)} = -{\alpha'\over2}\ln|z_i|\ ,\quad\quad V'_i(0) = z_i\ ,
\label{B0tau}
\eeq
and we find
\beq
\frac{\mbox{d}^2 z_i}{|V_i'(0)|^2} \ = \ \frac{\mbox{d}^2
z_i}{|z_i|^2}
\ \alimit \ \frac{4\pi}{\alpha'} \mbox{d} \tau^{(0)} \ , 
\eeq
where again the factor of $2\pi$ is obtained from the empty phase integral.

If instead $z_i$ is inserted on the $j$-th internal propagator,
$j=1,\ldots,h$, we have~\cite{RS}
\beq
\tau^{(j)} = -{\alpha'\over2}\ln\left|{z_i-\za{j}\over\za{j}}\right| =
-{\alpha'\over2}\ln\left|z_i-\za{j}\right| - \tau_{\alpha_j}   \ ,
\quad\quad V'_i(0) = |z_i -\za{j}| \ , 
\label{Bitau}
\eeq 
in which case
\beq
\frac{\mbox{d}^2 z_i}{|V_i'(0)|^2} \ = \ \frac{\mbox{d}^2
z_i}{|z_i-z_{\alpha_j}|^2} \ = \ \frac{\mbox{d}^2
(z_i-z_{\alpha_j})}{|z_i-z_{\alpha_j}|^2}
\ \alimit \ \frac{4\pi}{\alpha'} \mbox{d} (\tau^{(j)}+\tau_{\alpha_j}) \ =
\ \frac{4\pi}{\alpha'} \mbox{d} \tau^{(j)} \ . \label{partofmeasure}
\eeq
Here we effectively treated $\za{j}$ and $\tau_{\alpha_j}$ as
constants. For $j=1$ they actually are constants, since we have fixed
$\za{1}=1$, corresponding to $\tau_{\alpha_1}=0$. For $j=2,\ldots,h$ our procedure 
is justified by the fact that one should really
consider \eq{partofmeasure} as a part of the total integration measure
$\mbox{d}^2 {\cal M}$, and the changes of variables $(z_i,\za{j})
\rightarrow (z_i-\za{j},\za{j})$ and $(\tau^{(j)},\tau_{\alpha_j})
\rightarrow (\tau^{(j)} - \tau_{\alpha_j},\tau_{\alpha_j})$ have unit
Jacobian. 
 
In summary,
\beq
\mbox{d}^2{\cal M}_2\quad\alimit\quad \left({4\pi\over\alpha'}\right)^{N}
\prod_{\mu=0}^h \prod_{n=1}^{N_{\mu}} d\tau_n^{(\mu)} \ , \label{dmtwo}
\eeq
and by combining eqs.~\eq{dmone} and \eq{dmtwo}
the total measure is therefore reduced to 
\beqa
\lefteqn{\mbox{d}^2{\cal M}\, \alimit\, } \label{dm} \\
& & \left( {4\pi\over\alpha'}\right)^{3h+N} 
\mbox{d}T \left( \prod_{i=1}^h 
\mbox{d}\bar{T}_i \right) \mbox{d}\tau_{\alpha_1} \left( \prod_{i=2}^h
\mbox{d}\tau_{\alpha_i} \mbox{d}\tau_{\beta_i} \right) 
\exp\left[ {4\over\alpha'}(T+\sum_{i=1}^h{\bar T}_i)\right]\,
\prod_{\mu=0}^h \prod_{n=1}^{N_{\mu}} \mbox{d}\tau_n^{(\mu)} \ , \nn
\eeqa
after we integrate out the $3h+N$ phases.

\section{Summary}
\setcounter{equation}{0}
By combining eqs.~\eq{overallconstant}, \eq{exponent},
\eq{determinantformula} and \eq{dm} we find that in {\em any}
$\Phi^3$-like corner of moduli space that corresponds to a diagram of
the class ${\cal C}^{(h+1)}_{N_0,\ldots,N_h}$, the integrand of the
string formula
\eq{stramp} reduces in the $\alpha' \rightarrow 0$ and $g$ fixed limit to
the following expression
\begin{eqnarray}
& & {\cal N}' \
g^{N+2h} \mbox{d}T T^{-D/2}  
\ \cdot \prod_{i=1}^h \mbox{d} {\bar T}_i \cdot \mbox{d} \tau_{\beta_1}
\cdot \prod_{i=2}^h \mbox{d} \tau_{\alpha_i} \mbox{d}
\tau_{\beta_i}  \nonumber \\
& & \times 
\prod_{\mu=0}^h \prod_{n=1}^{N_{\mu}} \mbox{d}
\tau_n^{(\mu)} \cdot \  \ (\mbox{det}A)^{-D/2}\ 
\exp\left[\frac{4}{\alpha'} (T+\sum_{i=1}^h{\bar T}_i)\right]  \nn\\
& & \times \ \exp\left[{1\over2}\sum_{i,j=0}^h\sum_{n=1}^{N_i}
    \sum_{n'=1}^{N_j}p_n^{(i)} p_{n'}^{(j)} 
G_{ij}^{(h)}(\tau_n^{(i)},\tau_{n'}^{(j)})\right] \ ,
\label{finalexpression} 
\end{eqnarray}
where
\beq
{\cal N}' = \left( \frac{1}{4\pi} \right)^{D(h+1)/2} \ .
\eeq
If we make the substitution $4/\alpha' \rightarrow -M^2$ in the
exponent, the expression \eq{finalexpression} clearly reproduces the
integrand appearing in the particle formula \eq{ptamptwo}, except for
the unspecified overall normalization
${\cal N}$. 

Of course, in order to reproduce the particle formula completely one
should also consider the region of integration for the SPTs, not only
the SPT integrand.
It should be possible to {\em derive} the region of integration 
from string theory, but this is not
trivial. First of all it requires a careful analysis of the 
region of integration of the string moduli in
each $\Phi^3$-like corner of moduli space. The well-known mapping into
SPTs allows one to translate the region of integration for the string
moduli into a region of integration for the SPTs. Next, one should sum
the contributions from all the $\Phi^3$-like corners that correspond to 
diagrams in ${\cal C}^{(h+1)}_{N_0,\ldots,N_h}$ and one should analyze
how the SPT integrations of the various corners combine into a single
larger region of integration. But even at this point one cannot expect to 
recover the region of integration of the particle formula
\eq{ptamptwo}. The point is that the integral in the string formula 
\eq{stramp} is only over {\em one copy} of moduli space, whereas the
region of integration of the SPTs in the particle formula corresponds
to a region of integration for the string moduli that covers {\em
several} copies of moduli space. For example, the SPT configurations
related by an interchange of $\tau_{\alpha_i}$ and $\tau_{\beta_i}$
(for any $i=1,\ldots,h$) are both included in the SPT integral
\eq{ptamp} even though they certainly correspond to the same Riemann
surface, i.e. the same point in moduli space, inasmuch as they are
related by the modular transformation $(\za{i},\zb{i}) \rightarrow
(\zb{i},\za{i})$ that replaces the $i$'th Schottky generator by its inverse.
These questions are obviously of importance for any practical
application of the string multiloop techniques, not only to $\Phi^3$-theory but
also to Yang-Mills theory, and we hope to address them in a future publication.

In conclusion, we have argued in this paper that the integral over
moduli space defining the $N$-tachyon $(h+1)$-loop amplitude in the
closed bosonic string reduces in the $\alpha' \rightarrow 0$ limit
(for fixed tachyon coupling constant) to a sum of contributions from
all $\phi^3$-like corners of moduli space, with each contribution
being in the form of a SPT integral. We have explicitly shown how the
modular integrand of the string 
formula \eq{stramp} reduces to the SPT integrand of the $\Phi^3$ particle
theory formula \eq{ptamptwo} in
any corner corresponding to a $\Phi^3$-diagram belonging to the class
${\cal C}_{N_0,N_1,\ldots, N_h}$.

The main simplifying feature of the $\Phi^3$-theory toy model was the
absence of contact terms, which was reflected by the fact that the
string modular integrand $\mbox{d}^2 {\cal M}$ had to scale like 
$(\alpha')^{-(N+3h)}$,
q.v. eq.~\eq{measurescale}. In the
physically more interesting case of the $N$-gluon $(h+1)$-loop
amplitude, where one takes $\alpha' \rightarrow 0$ for fixed
Yang-Mills coupling constant, the $\alpha'$ power counting is
different and surviving contributions in the $\alpha' \rightarrow 0$ 
limit are no longer compelled to appear only from $\Phi^3$-like
corners of moduli space. There will also be other contributions. 
At the one-loop level these extra contributions (called Type II) were
studied in great detail in ref.~\cite{PDVBIG}. They may be removed by
an integration-by-parts procedure (as shown in ref.~\cite{BKibp}) but
this does not seem possible at more than one loop.
Recent work by Magnea
and Russo~\cite{MagneaRusso} 
in the specific case of the two-loop vacuum amplitude
constitutes a successful first step towards a full understanding of
these extra contributions at the multiloop level.

\appendix

\section*{Appendix A. The symmetric parametrization of the two-loop case}
\setcounter{equation}{0}
\setcounter{section}{1}
In this appendix we consider the special case of two loops $(h=1)$.
The unique Schmidt-Schubert type vacuum diagram consists of three
propagators, joined together at an ``upper'' and a ``lower'' 3-point
vertex, and there exists a symmetric world-line
parametrization~\cite{SSphi,HTS} which
clearly displays the symmetries of this diagram: One defines a SPT
$\tau^{(i)}_{\rm sym}$, $i=1,2,3$, along each of the propagators,
taking $\tau^{(i)}_{\rm sym} = 0$ at the ``upper'' vertex and
$\tau^{(i)}_{\rm sym} = t_i$ at the ``lower'' vertex (see Fig. 2).
\begin{figure}
\begin{center}
\input{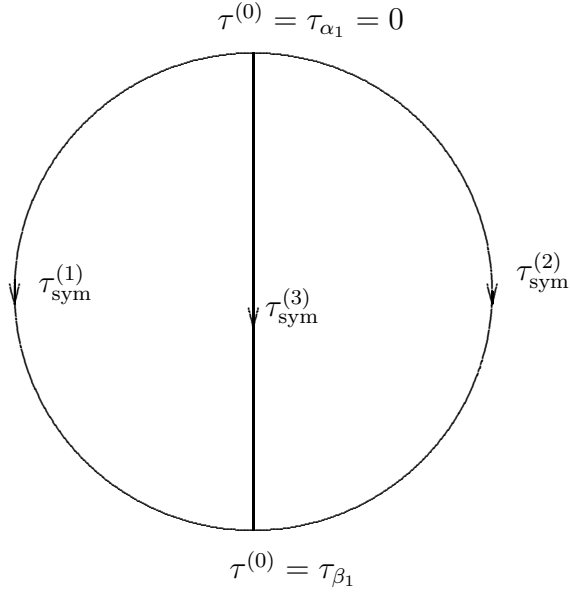}
\end{center}
\caption{The symmetric SPT parametrization of the two-loop 
Schmidt-Schubert vacuum diagram.}
\end{figure}

The relation between the symmetric parametrization and the one
underlying eq.~\eq{ptamptwo} is given by
\beqa
t_1 & = & \tau_{\beta_1}  \label{relation} \\
t_2 & = & T - \tau_{\beta_1} \nn \\
t_3 & = & \bar{T}_1 \nn \ , 
\eeqa
where we used translational invariance along the fundamental loop to
choose the ``upper'' vertex to be at $\tau_{\alpha_1} = 0$.

In the symmetric parametrization a leg on the fundamental loop with
SPT $\tau^{(0)}$ is considered to sit on the first
propagator, with $\tau^{(1)}_{\rm sym} = \tau^{(0)}$,
if $\tau^{(0)} \in [0;\tau_{\beta_1}]$, and to sit on the second
propagator, with $\tau^{(2)}_{\rm sym} = T-\tau^{(0)}= t_1 + t_2 -
\tau^{(0)}$, if $\tau^{(0)} \in [\tau_{\beta_1};T]$.
Whereas along the internal propagator we simply have
$\tau^{(3)}_{\rm sym} = \tau^{(1)}$.

If we rewrite eq.~\eq{ptamptwo} for $h=1$ in terms of the symmetric
parametrization and restrict ourselves to the contribution where $N_i$
legs are inserted on the $i$'th propagator, $i=1,2,3$, we
obtain~\cite{HTS}
\beqa
\lefteqn{
\Gamma_{N_1,N_2,N_3} \ = \ {\cal N} g^{N+2} \cdot
\prod_{a=1}^3\int_0^{\infty} \mbox{d}t_a e^{-M^2 t_a} \cdot 
(t_1t_2+t_2t_3+t_3t_1)^{-D/2} } \label{another} \\
& &  \prod_{n=1}^{N_1} \int_0^{t_1} \mbox{d}\tau^{(1)}_n \cdot
     \prod_{n=1}^{N_2} \int_0^{t_2} \mbox{d}\tau^{(2)}_n \cdot
     \prod_{n=1}^{N_3} \int_0^{t_3} \mbox{d}\tau^{(3)}_n \nn \\
& &  \times \exp[{1\over2}\sum_{a=1}^3 \sum_{j,k=1}^{N_a} p^{(a)}_j p^{(a)}_k 
                      G^{\rm sym}_{aa}(\tau^{(a)}_j,\tau^{(a)}_k)
     +\sum_{a=1}^3 \sum_{j=1}^{N_a}\sum_{k=1}^{N_{a+1}} p^{(a)}_j p^{(a+1)}_k 
      G^{\rm sym}_{aa+1}(\tau^{(a)}_j,\tau^{(a+1)}_k) ], \nn
\eeqa
where we adopted a cyclic notation, e.g. $t_{a+3} = t_a$ and dropped the
label ``sym'' on the SPTs. The symmetric Green functions appearing in
eq.~\eq{another}, 
\beq
G^{\rm sym}_{aa}(x,y)=\vert x-y\vert -{t_{a+1}+t_{a+2}\over 
                   t_1t_2+t_2t_3+t_3t_1}(x-y)^2\,,       \label{gaa}
\eeq
\beq
G^{\rm sym}_{aa+1}(x,y)=x+y-{x^2t_{a+1}+y^2t_{a}+(x+y)^2t_{a+2} \over
                      t_1t_2+t_2t_3+t_3t_1}\,,          \label{gaa1}
\eeq
can be obtained from the Green functions $G_{00}^{(1)}$ and
$G_{10}^{(1)}$ by making the appropriate changes of SPT variables.
In view of this it 
should be clear that the string theory derivation of the Green
functions \eq{gh00}, \eq{ghii}, \eq{ghi0} and \eq{ghij} given in
ref.~\cite{RS} include also the symmetric Green functions \eq{gaa} and
\eq{gaa1}. 

Indeed, we can obtain the Green functions \eq{gaa} and \eq{gaa1}
directly from the string theory Green function as follows:

First we define the SPTs pertaining to the fundamental loop
\beq
T=-{\alpha'\over2}\ln\vert k_0\vert, \quad
\tau_{\alpha_1}=-{\alpha'\over2}\ln\vert \za{1} \vert, \quad 
\tau_{\beta_1}=-{\alpha'\over2}\ln\vert \zb{1} \vert, 
\eeq
and then we transform these according to \cite{HTS}, q.v. eq.~\eq{relation}
above
\beqa
t_2 &=& T-|\tau_{\alpha_1}-\tau_{\beta_1}| = -{\alpha'\over2}\ln\left| k_0
      {(\za{1}-\zb{1})^2\over \za{1}\zb{1}} \right|, \\
t_1 &=& |\tau_{\alpha_1}-\tau_{\beta_1}| = {\alpha'\over2}
        \ln\left|{(\za{1} -\zb{1})^2\over\za{1}\zb{1}}\right|.
\eeqa
The proper time for the internal propagater is given by
(q.v. eq.~\eq{Tintprop}) 
\beq
t_3 = -{\alpha'\over2}\ln\left| k_1{(\za{1}-\zb{1})^2\over 
               \za{1}\zb{1}} \right|.
\eeq
We may use symmetry of the world-sheet Green function under the
exchange of $z_{\alpha_1}$ and $z_{\beta_1}$ to arrange for
$\tau_{\beta_1} > \tau_{\alpha_1}$ or in other words, $|z_{\beta_1}|
\ll |z_{\alpha_1}|$. Then, in terms of
the parameters $t_1$, $t_2$ and $t_3$ the period matrix \eq{period}
becomes
\beq
(2\pi\mbox{Im}\tau)_{\mu\nu}={2\over\alpha'}
\left(\begin{array}{ll} t_1+t_2 &\quad t_1\\
                           t_1  &\quad t_1+t_3 \end{array}\right) \, ,
\label{periodtwo}
\eeq
and the world-sheet Green function \eq{sgreen} assumes the form
\beqa
G^{(1)}_{\rm str}(z_1,z_2) & = &  \ln \left| 
\frac{z_1 - z_2}{(V_1'(0))^{1/2} (V_2'(0))^{1/2}} \right| 
-{\alpha'\over4}(t_1t_2+t_2t_3+t_3t_1)^{-1}           \label{gstr} \\
& & \times [ (t_1+t_3)\Omega_0^2 
      - 2t_1\Omega_0\Omega_1
            +(t_1+t_2)\Omega_1^2 ] \ .        \nn 
\eeqa
We notice how the determinant of the period matrix \eq{periodtwo} gives
rise to the factor 
$t_1t_2+t_2t_3+t_3t_1$ appearing in the worldline integral \eq{another}. 
To recover the various Green functions \eq{gaa} and \eq{gaa1}  
from \eq{gstr} we have to specify the appropriate pinching
limits. This can be done by following the rules given in ref.~\cite{RS}.

If $z_i$ describes a leg inserted on the internal propagator, we may
consider the pinching limit $|z_i - z_{\alpha_1}| \ll |z_{\alpha_1}|$
and the corresponding SPT is (q.v. eq.~\eq{Bitau})
\beq
\tau^{(3)}_i = 
     -{\alpha'\over2}\ln\left|{z_i-\za{1}\over\za{1}}\right| \ . 
\eeq
If instead $z_i$ describes an external leg located on the 
fundamental loop, we have
\beqa
\tau^{(1)}_i & = & -{\alpha'\over2}\ln\left|{z_i\over \za{1}}\right| \quad
\mbox{for} \quad |\zb{1}| \ll |z_i| \ll |\za{1}| \ ,  \\
\tau^{(2)}_i & = & -{\alpha'\over2}\ln\left|k_0 {\za{1} \over z_i
}\right| \quad
\mbox{for} \quad |k_0 \za{1}| \ll |z_i| \ll |\zb{1}| \ .   \nn  
\eeqa                                                            
It is sufficient to consider 
the two cases $G^{\rm sym}_{31}$ and $G^{\rm sym}_{33}$, 
because of cyclic symmetry. 

\begin{figure}
\begin{center}
\input{ampfig3}
\end{center}
\caption{The $\Phi^3$ diagram relevant for the Green function 
$G_{31}^{\rm sym}$. }
\end{figure}
In the case of $G^{\rm sym}_{31}(\tau^{(3)},\tau^{(1)})$ we have
the diagram shown in Fig.~3, which according to the rules of
ref.~\cite{RS} corresponds to the pinching limit
\beq
|\zb{1}| \ll |z_2| \ll |\za{1}| \simeq |z_1| 
\quad\mbox{and}\quad |z_1 - \za{1}| 
 \ll |\za{1}| \ , \label{pinchingone}
\eeq
and the choice of local coordinates
\beq     
\left\{ \begin{array}{l}
V_1'(0)=\vert z_1 - \za{1} \vert \\
V_2'(0)=\vert z_2 \vert\, . \\
\end{array}\right.                             \label{vzero31}
\eeq      
In the pinching limit \eq{pinchingone} we find
\beqa
\Omega_0&=&\ln\left|{z_2\over z_1}\right| \ \simeq \ \ln
\left|{z_2\over \za{1}} \right| \ = \ 
         -{2\over\alpha'} \tau^{(1)}\ , \label{omeganul} \\
\Omega_1&=& \ln \left| {\zb{1}-z_2 \over \zb{1} - z_1} {\za{1}-z_1
\over \za{1} - z_2} \right| \ \simeq \
\ln\left|{z_2 (\za{1}-z_1) \over (\za{1})^2 } \right| \ = \ 
-{2\over\alpha'}(\tau^{(3)}+ \tau^{(1)})  , \label{omegaet}
\eeqa
\beq
\ln \left| 
\frac{z_1 - z_2}{(V_1'(0))^{1/2} (V_2'(0))^{1/2}} \right| \ \simeq \
{1 \over 2} \ln \left| {(\za{1})^2 \over z_2 (z_1-\za{1})} \right| \ =
\ {1\over\alpha'}(\tau^{(1)} + \tau^{(3)})\, , \label{treepart}
\eeq
and thus, by inserting \eq{omeganul}, \eq{omegaet} and \eq{treepart}
into eq.~\eq{gstr},
\beq
G^{(1)}_{\rm str}(z_1,z_2) \quad\alimit\quad 
{1\over\alpha'}G^{\rm sym}_{31}(\tau^{(3)},\tau^{(1)})\,.
\eeq
\begin{figure}
\begin{center}
\input{ampfig4}
\end{center}
\caption{The $\Phi^3$ diagram relevant for the Green function 
$G_{33}^{\rm sym}$. }
\end{figure}
In the case of $G^{\rm sym}_{33}(\tau^{(3)}_1,\tau^{(3)}_2)$ we 
consider instead the pinching limit (q.v. Fig.4)
\beq
|z_2-\za{1}| , |z_1-\za{1}| \ll |\za{1}| \quad \mbox{and} \quad
|\zb{1}| \ll |\za{1}| \ ,
\eeq 
with
\beq
\left\{ \begin{array}{l}
V_1'(0)=\vert z_1 - \za{1} \vert \\
V_2'(0)=\vert z_2 - \za{1} \vert\,, \\
\end{array}\right.                               \label{vzero33}
\eeq                   
and we find
\beqa
\Omega_0 &\simeq& 0\,, \\
\Omega_1 &\simeq& \ln\left|{z_1-\za{1} \over z_2-\za{1}}\right| 
\ = \ -{2\over\alpha'}(\tau^{(3)}_1 -\tau^{(3)}_2)\,,
\eeqa
\beq
\ln \left| 
\frac{z_1 - z_2}{(V_1'(0))^{1/2} (V_2'(0))^{1/2}} \right|\quad 
\simeq\quad {1\over\alpha'}|\tau^{(3)}_1 - \tau^{(3)}_2| \,,
\eeq
and thus 
\beq
G^{(1)}_{\rm str}(z_1,z_2) \quad\alimit\quad 
{1\over\alpha'}G^{\rm sym}_{33}(\tau^{(3)}_1,\tau^{(3)}_2)\,.
\eeq
It is now easy to see that if we evaluate the two-loop version of the
string ``master formula'' \eq{stramp} in the
relevant pinching limits, mapping the string moduli directly into the
SPTs of the symmetric parametrization, we recover the entire SPT integrand  
of the particle formula \eq{another}, except for the overall numerical
constant. 

In conclusion, we have shown explicitly how the procedure for
extracting $\Phi^3$ particle theory from the bosonic string leads to
the correct two-loop particle formulae, regardless of which of the two
SPT parametrizations we choose. This should in fact be clear {\it a priori},
since the process by which we extract contributions from the
$\Phi^3$-like corners of moduli space, including the choice of local
coordinates, is entirely geometrical and does not rely on any
specific choice of SPT parametrization.

\section*{Appendix B. Translational invariance along the fundamental loop}
\setcounter{equation}{0}
\setcounter{section}{2}
In this appendix we show how to obtain the worldline formula
\eq{ptamptwo} from the more standard expression \eq{ptamp} by using
the invariance of the integrand under the transformation
\beq
\begin{array}{lll}
\left. {\begin{array}{lll} 
\tau_{\alpha_i} & \rightarrow & \tau_{\alpha_i} + c \\
\tau_{\beta_i} & \rightarrow & \tau_{\beta_i} + c
\end{array}} \right\} & \mbox{for} & i =1,\ldots, h \\
\ \, \tau_n^{(0)} \ \rightarrow \ \tau_n^{(0)} + c & \mbox{for} &
n=1,\ldots,N_0 \end{array} \label{shift}
\eeq
that translates all SPTs pertaining to the fundamental loop by the
same constant, $c$.

This invariance follows from the similar property of the bosonic Green
function \eq{GB},
\beq
G_B(\tau_a,\tau_b) = G_B(\tau_a+c,\tau_b+c) \ ,
\eeq
once we notice that the SPTs of the fundamental loop enter
eq.~\eq{ptamp} {\em only} as arguments of $G_B$ and that {\em every}
$G_B$ appearing has two SPTs of the fundamental loop as arguments.

If we introduce a simplified notation, where $x_0,x_1,\ldots,x_n$
denote the $n+1=2h+N_0$ SPTs of the fundamental loop, and the integrand
of the particle formula \eq{ptamp} is called $f(x_0,x_1,\ldots,x_n)$,
then the invariance \eq{shift} allows us to rewrite the amplitude
\eq{ptamp} as
\beqa
{\cal A} & = & \int_0^T \mbox{d} x_0 \int_0^T \mbox{d} x_1 \ldots
\int_0^T \mbox{d} x_n \ f(x_0,x_1,\ldots,x_n) \\
& = & \int_0^T \mbox{d} x_0 \int_0^T \mbox{d} x_1 \ldots
\int_0^T \mbox{d} x_n \ f(0,x_1-x_0,\ldots,x_n-x_0) \nn \\
& = & \int_0^T \mbox{d} x_0 \int_{-x_0}^{-x_0+T} \mbox{d} x_1 \ldots
\int_{-x_0}^{-x_0+T} \mbox{d} x_n \ f(0,x_1,\ldots,x_n) \ . \nn
\eeqa
At this point the dependence on $x_0$ has disappeared from the
integrand but remains in the limits of integration.

If $f$ was periodic with period $T$ in each variable, we could just
replace the integration region
\beq
\int_{-x_0}^{-x_0+T} \mbox{d}x_i \rightarrow \int_{0}^{T} \mbox{d}x_i
\label{replace}
\eeq
for each $i=1,\ldots,n$. But things are not so straightforward,
because the Green function $G_B$, from which $f$ is constructed, is
not periodic. Rather, it satisfies
\beqa
G_B(\tau_1+T,\tau_2) & = & G_B(\tau_1,\tau_2) \quad \mbox{if} \quad
\tau_1 < \tau_2 \label{periodicitya} \\
G_B(\tau_1-T,\tau_2) & = & G_B(\tau_1,\tau_2) \quad \mbox{if} \quad
\tau_1 > \tau_2 \ , \label{periodicityb}
\eeqa
provided that $|\tau_1-\tau_2| < T$.

Fortunately, this ``restricted periodicity'' turns out to be
sufficient to justify the shift \eq{replace} of the integration
region.

To demonstrate the underlying mechanism, consider first the case $n=1$
(the two-loop vacuum diagram). Then we have
\beqa
{\cal A} & = & \int_0^T \mbox{d}x_0 \int_{-x_0}^{-x_0+T}
\mbox{d} x_1 \ f(0,x_1)  \\
& = & \int_0^T \mbox{d}x_0  \left( \int_{-x_0}^0 \mbox{d} x_1 \, f(0,x_1) +
\int_{0}^{-x_0+T} \mbox{d} x_1 \, f(0,x_1) \right) \ . \nn 
\eeqa
In the first term, $x_1$ is the smallest of the two SPT arguments, 
since $x_1<0$ and the other is zero. 
In view of eq.~\eq{periodicitya} we may then write
\beq
f(0,x_1) = f(0,x_1+T) \ ,
\eeq
inasmuch as $f$ is constructed entirely from $G_B$. If we then change
integration variables from $x_1$ to $x_1' = x_1+T$ we obtain
\beqa
{\cal A} & = & \int_0^T \mbox{d}x_0  
\left( \int_{-x_0+T}^T \mbox{d} x_1' f(0,x_1') +
\int_{0}^{-x_0+T} \mbox{d} x_1 f(0,x_1) \right) \\
& = & \int_0^T \mbox{d}x_0 \int_{0}^{T}
\mbox{d} x_1 f(0,x_1) \ = \  T \int_0^T \mbox{d}x_1 f(0,x_1) \ . \nn
\eeqa
In the general case we break each of the $n$ $x_i$ integrations into
two regions,
\beq
\int_{-x_0}^{-x_0+T} \mbox{d}x_i \ = \ \int_{-x_0}^{0} \mbox{d}x_i + 
\int_0^{-x_0+T} \mbox{d}x_i \ . \label{startpoint}
\eeq
Then ${\cal A}$ becomes a sum of $2^n$ terms, where in a ``typical''
term we have $n_1$ integrations that run over negative values (from
$-x_0$ to $0$) and $n-n_1$ integrations that run over positive values
(from $0$ to $-x_0+T$), $n_1=0,1,\ldots,n$.

In each ``typical'' term we may break the integration region for the
$n_1$ variables that run over negative values into $n_1!$ smaller
integration regions, each corresponding to a possible ordering of the
$n_1$ negative variables.

Let us assume for notational simplicity that the $n_1$ negative
variables are $x_1,\ldots,x_{n_1}$. Then we rewrite the ``typical''
term as
\beqa
& \displaystyle{ \int_0^T \mbox{d} x_0 \int_{-x_0}^0 \mbox{d} x_1 \ldots 
\int_{-x_0}^0 \mbox{d} x_{n_1} \int_0^{-x_0+T} \mbox{d} x_{n_1+1}
\ldots \int_0^{-x_0+T} \mbox{d} x_{n} \ f(0,x_1,\ldots,x_n) \ = } &
\nn \\
& \displaystyle{
\sum_{\sigma \in S_{n_1}} \int_0^T \mbox{d} x_0 \int_{-x_0}^0
\mbox{d} x_{\sigma(1)} \int_{x_{\sigma(1)}}^0
\mbox{d} x_{\sigma(2)} \ldots \int_{x_{\sigma(n_1-1)}}^0
\mbox{d} x_{\sigma(n_1)}  } & \nn \\
& \displaystyle{ \times \int_0^{-x_0+T} \mbox{d} x_{n_1+1}
\ldots \int_0^{-x_0+T} \mbox{d} x_{n} \ f(0,x_1,\ldots,x_n) \ , } &
\label{typterm}
\eeqa
where the sum is over all permutations of the integers
$\{1,2,\ldots,n_1\}$. Since
\beq
x_{\sigma(1)} < x_{\sigma(2)} < \ldots < x_{\sigma(n_1)} < 0 <
x_{n_1+1}, \ldots , x_n \ , 
\eeq
we may use eq.~\eq{periodicitya} to replace $x_{\sigma(1)}$ by
$x_{\sigma(1)}+T=x'_{\sigma(1)}$ in the integrand $f$ and then change
variables to $x'_{\sigma(1)}$. Since $x'_{\sigma(1)}$ runs from
$-x_0+T$ to $T$, i.e. is positive, it is now $x_{\sigma(2)}$ which is
the smallest SPT variable, and we may repeat the above step. By
continued use of this trick, the ``typical'' term \eq{typterm} becomes
\beqa
\lefteqn{  \sum_{\sigma \in S_{n_1}} \int_0^T \mbox{d} x_0 \int_{-x_0+T}^T
\mbox{d} x'_{\sigma(1)} \int_{x'_{\sigma(1)}}^T
\mbox{d} x'_{\sigma(2)} \ldots \int_{x'_{\sigma(n_1-1)}}^T
\mbox{d} x'_{\sigma(n_1)} } \ \\
& & \ \ \ \ \times \int_0^{-x_0+T} \mbox{d} x_{n_1+1}
\ldots \int_0^{-x_0+T} \mbox{d} x_{n} \
f(0,x_1',\ldots,x_{n_1}',x_{n_1+1}, \ldots , x_n )  \nn \\
& & = \ \int_0^T \mbox{d}x_0 \int_{-x_0+T}^T \mbox{d} x_1' \ldots  
\int_{-x_0+T}^T \mbox{d} x_{n_1}' \nn \\
& & \ \ \ \ \times \int_0^{-x_0+T} \mbox{d}x_{n_1+1} 
\int_0^{-x_0+T} \mbox{d}x_{n} \ f(0,x_1', \ldots , x_{n_1}',x_{n_1+1},
\ldots , x_n ) \ . \nn 
\eeqa
Effectively, all integrations over negative SPTs have been shifted to
positive values and when all the $2^n$ ``typical'' terms are
recombined, the $x_i$ integration now becomes
\beq
\int_{-x_0+T}^T \mbox{d} x_i \ + \ \int_0^{-x_0+T} \mbox{d} x_i \ = \
\int_0^T \mbox{d} x_i \ .
\eeq
In conclusion, we arrive at the expression \eq{ptamptwo} for the
amplitude,
\beqa
{\cal A} & = & \int_0^T \mbox{d} x_0 \int_0^T \mbox{d} x_1 \ldots
\int_0^T \mbox{d} x_n \ f(0,x_1,\ldots,x_n) \\
& = & T \int_0^T \mbox{d} x_1 \ldots
\int_0^T \mbox{d} x_n \ f(0,x_1,\ldots,x_n) \ . \nn
\eeqa
%


%
\end{document}